\documentclass{pasa}%

\usepackage{graphicx}
\usepackage[natbib, style=apa, mincitenames=3, maxcitenames=3]{biblatex}
\title[Orbit determination using the MWA]{Demonstration of Orbit Determination for LEO Objects using the Murchison Widefield Array}

\author[Prabu et al.]{Prabu, S.$^{1,4}$, Hancock, P.$^2$, Zhang, X.$^{3,5}$, and Tingay, S.J.$^{1}$
\affil{$^1$International Centre for Radio Astronomy Research, Curtin University, Bentley, WA 6102, Australia}%
\affil{$^2$Curtin Institute for Computation, Curtin University, GPO Box U1987, Perth, 6845, WA, Australia}%
\affil{$^3$CSIRO Space and Astronomy, 26 Dick Perry Avenue, Kensington, WA 6151, Australia}
\affil{$^4$CSIRO Space and Astronomy, Corner Vimiera \& Pembroke Roads, Marsfield, NSW 2122, Australia}
\affil{$^5$LESIA, Observatoire de Paris, Université PSL, CNRS, Sorbonne Université, Université Paris Cité, 5 place Jules Janssen, 92195 Meudon, France}
}%

\jid{PASA}
\doi{10.1017/pas.\the\year.xxx}
\jyear{\the\year}

\usepackage{aas_macros}
\usepackage{hyperref} 
\usepackage{lineno, blindtext}
\hypersetup{colorlinks,citecolor=blue,linkcolor=blue,urlcolor=blue}

\begin{document}

\begin{frontmatter}
\maketitle

\begin{abstract}
The rapidly increasing number of satellites in Earth's orbit motivates the development of Space Domain Awareness (SDA) capabilities using wide field-of-view sensor systems that can perform simultaneous detections. This work demonstrates preliminary orbit determination capability for Low Earth Orbit objects using the $36^{\circ}\times36^{\circ}$ field-of-view of the Murchison Widefield Array (MWA) at commercial Frequency Modulated (FM) frequencies (transmitters in $88-108$\, MHz range). Non-coherent passive radar techniques with the MWA produce spatially smeared detections, due to time averaging in the MWA's standard signal processing chain. The work develops methods to extract time-stamped measurements of a satellite's angular coordinates from these data. The developed method was tested on observations of $32$ satellite passes and the extracted measurements were used to perform orbit determination for the targets using a least-squares fitting approach. The target satellites span a range in altitude and Radar Cross Section, providing examples of both high and low signal-to-noise detections.  The estimated orbital elements for the satellites are validated against the publicly available Two Line Element (TLE) updates provided by the Space Surveillance Network (SSN) and the preliminary estimates are found to be in close agreement. The work also tests for re-acquisition for one target using the  orbital elements and finds the trajectory predicted by the method to coincide within $0.2^{\circ}$ cross-track and $0.3^{\circ}$ in-track for a subsequent pass, reduced to approximately $0.1^{\circ}$ cross-track (less than one kilometre) if two passes are used to predict the subsequent pass (using simple two-body propagation). The median uncertainty in the angular position for objects in LEO (range less than 1000 km) is found to be $860$ m in the cross-track direction and $780$ m in the in-track direction, which are comparable to the typical uncertainty of $\sim$1000 m in the publicly available TLE information. The techniques, therefore, demonstrate the MWA to be capable of being a valuable contributor to the global SDA community. Based on the understanding of the MWA SDA system, this paper also briefly describes methods to mitigate the impact of FM-reflecting LEO satellites on radio astronomy observations, and how maintaining a catalog of FM-reflecting LEO objects is in the best interests of both SDA and radio astronomy.

\end{abstract}

\end{frontmatter}

\section{INTRODUCTION }

\label{sec:intro}

For practical reasons the motions of a human-made object around the Earth are often approximated in terms of an idealised two body system that can be defined using six Keplerian/orbital elements. However, the approximation is only valid near the epoch at which measurements are made, as they change due to atmospheric drag, J2 effects \citep{mishne2004formation} (perturbations in orbit caused due to the Earth not being a perfect sphere), Solar Radiation pressure, and gravitational effects of the Sun and Moon. Hence, orbital elements must be updated  in order to maintain a current and accurate understanding of the state of the near Earth space environment (Space Domain Awareness: SDA).

With the ongoing increase in the number of satellites in Earth orbit, catalog maintenance can be a challenging task. Thus multiple sensors are required to work together to perform detections at high rates, and maintain catalogs through data fusion. Some existing SDA sensors already work in pairs \citep{cordelli2019use}, consisting of one wide field-of-view (FOV) system that performs the detection of the objects along with preliminary orbit estimates, followed by precise orbit determination using higher resolution measurements obtained from instruments that often have a smaller FOV. 

In this paper, a preliminary orbit estimation capability is demonstrated for Low Earth Orbit (LEO) objects using the Murchison Widefield Array (MWA) \citep{Tingay2013TheFrequencies}, using non-coherent passive radar techniques \citep{Tingay2013OnFeasibility, zhang2018limits, prabu_dev, prabu_survey} over very large FOVs at commercial Frequency Modulated (FM) frequencies ($\sim$88 MHz to $\sim$108 MHz in Australia). Radio telescopes such as LOw Frequency ARray (LOFAR) \citep{klos2020possibility, klos2021experimental} and the Italian Northern Cross telescope array \citep{2022AcAau.198..111M} have also been used to perform SDA observations, and recent studies \citep{mcdowell2020low,gallozzi2020concerns,hainaut2020impact,mallama2020flat} have highlighted  the importance of understanding the near-Earth space environment for astronomy in optical, infrared, and radio wavebands. Developing and maintaining a catalog of LEO objects known to reflect FM signals can help understand the sources of Radio Frequency Interference (RFI) in the environment of the Murchison Radio Observatory (MRO) \citep{2020PASA...37...39T}, home to the MWA and the future low-frequency element of the Square Kilometre Array (SKA).

The MWA can perform SDA observations via the search for reflections of terrestrial FM signals from satellites using standard radio astronomy imaging techniques, known as  non-coherent passive radar. The detected signals are  smeared, typically over a couple of degrees, mainly due to the motion of the satellites during the image correlation time, as demonstrated in the previous work \citep{Tingay2013OnFeasibility,prabu_dev, prabu_survey}. The $\sim$0.5-2 second observation periods are driven by the MWA's standard configuration for its primary astrophysics mission. A detailed discussion of the smallest Radar Cross-Section (RCS) detected using the MWA can be found in \cite{prabu_survey} and \cite{PRABU2022}.

\begin{figure*}[h!]
\begin{center}
\includegraphics[width=\linewidth,keepaspectratio]{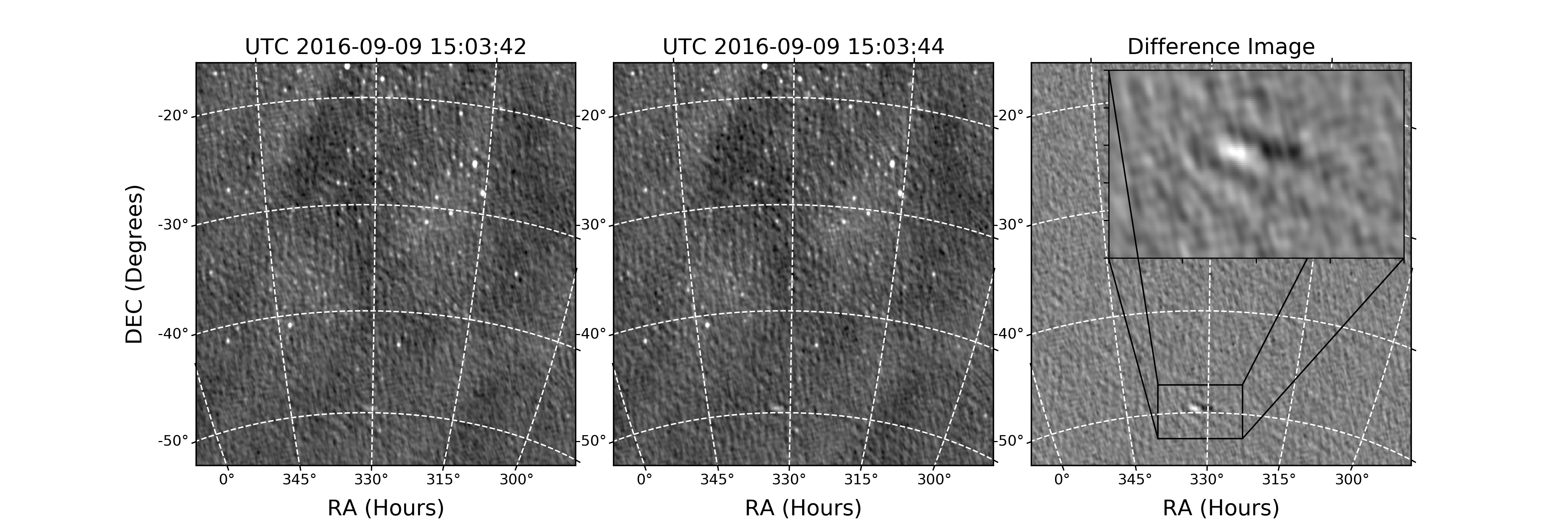}
\caption{The left and middle panels are consecutive $2\,$s images of the sky as observed by the MWA, and the right panel illustrates how difference images can be used to isolate transient events from static background sources, as well as stationary imaging artefacts.}
\label{Fig1DiffImg}
\end{center}
\end{figure*}

Deriving accurate angular position measurements and assigning time-stamps for these measurements can be a challenging task, due to these smearing effects. This work describes the methods to extract satellite angular position measurements from the spatially and temporally smeared data, and discuss how they can be used to estimate orbital elements for the detected objects.  The methods are tested on MWA observations of 32 LEO objects of varying altitude, inclination, RCS, and SNR reported in the previous work \citep{prabu_survey}. The satellite re-acquisition capability is tested using predictions based on the methods, for multiple observations of the Hubble Space Telescope (HST) during different orbits. 

The motivation for demonstrating a orbit estimation capability with the MWA is two-fold. Firstly, being able to perform valuable SDA measurements using the MWA is an added output of the versatile MWA FM band observations performed for radio-astronomy purposes. As demonstrated in this work using archived MWA observations, any MWA FM band observations can be re-used for LEO orbit estimation, and does not require any SDA specific scheduling of observations, and hence is a cost-effective addition to the existing global SDA effort. Second, its large FOV and 24/7 operational capability enables it to re-acquire LEO targets with ease. The orbital elements of LEO objects can vary significantly within a few hours, and the MWA's FOV and ability to continuously monitor the LEO environment potentially makes it an important SDA sensor in the age of satellite mega-constellations.

The non-coherent passive radar techniques are described in Section \ref{sec:background}. The methods developed to obtain satellite angular position measurements and how they can be used to perform orbit determinations are discussed in Section \ref{sec:methods}. The results from Section \ref{sec:results} are discussed in Section \ref{sec:discussion}, followed by a brief conclusion of the work in Section \ref{sec:conclusion}.
 
\section{SPACE SURVEILLANCE USING THE MURCHISON WIDEFIELD ARRAY}
\label{sec:background}

The MWA \citep{Tingay2013TheFrequencies} is a radio interferometer built as a precursor to the low frequency component of the SKA with a $36^{\circ}\times 36^{\circ}$ FoV. The MWA has 128 elements (each containing 16 dual polarised bow-tie antennas) that observe the radio sky from $70 - 300$\, MHz, and has an instantaneous bandwidth of $30.72$\,MHz. The MWA was designed for studying astronomical sources \citep{2013PASA...30...31B, 2019arXiv191002895B} but has also been demonstrated to be a novel instrument for performing SDA observations \citep{Tingay2013OnFeasibility,7944483}. Two different passive radar techniques have been developed for the MWA, namely coherent detection \citep{7944483,8835821,hennessy2020orbit} and  non-coherent detection  \citep{Tingay2013OnFeasibility,zhang2018limits,prabu_dev, prabu_survey}, and this work focuses on performing orbit determination using the data obtained from the non-coherent method.

The non-coherent detection method uses difference imaging on adjacent $2\,$s images to isolate transient satellite signals from static background astronomical emission \citep{zhang2018limits}. An example illustration of the difference imaging technique is shown in Figure \ref{Fig1DiffImg}. The left and middle panels of Figure \ref{Fig1DiffImg} shows two consecutive $2$\,s images of the MWA sky, and the difference image (right panel of Figure \ref{Fig1DiffImg}) separates transient events from the static background image. Satellite signals often appear as streaks in the difference images, with a positive head and the negative tail due to the satellite signal being present in both the individual frames. To increase the sensitivity of the search (as the signal of interest is narrow band), the MWA's $30.72\,$MHz instantaneous bandwidth is split into 768 fine channels (each of $40$\,kHz bandwidth) and search for events with peak flux density greater than 6 times the noise in the difference images created for each of these fine channels at every time-step. A detailed description of the steps and pipeline can be found in \citet{prabu_dev,prabu_survey} and the reader is referred to these papers for further details.

The detection of 74 objects are reported in \citet{prabu_survey}.  From these detections 32 satellite passes have been selected, of varying characteristics (Table \ref{tab1}) in order to test and demonstrate orbit determination, using the information contained in the detections. The selected 32 passes have higher SNR detections than the remainder passes not used in this work. Using low SNR events to perform angular position measurements often give large uncertainties, and hence is not considered in this work.

\section{METHODS:ANGULAR COORDINATE MEASUREMENTS}
\label{sec:methods}

\begin{figure*}[h!]
\begin{center}
\includegraphics[width=\linewidth,keepaspectratio]{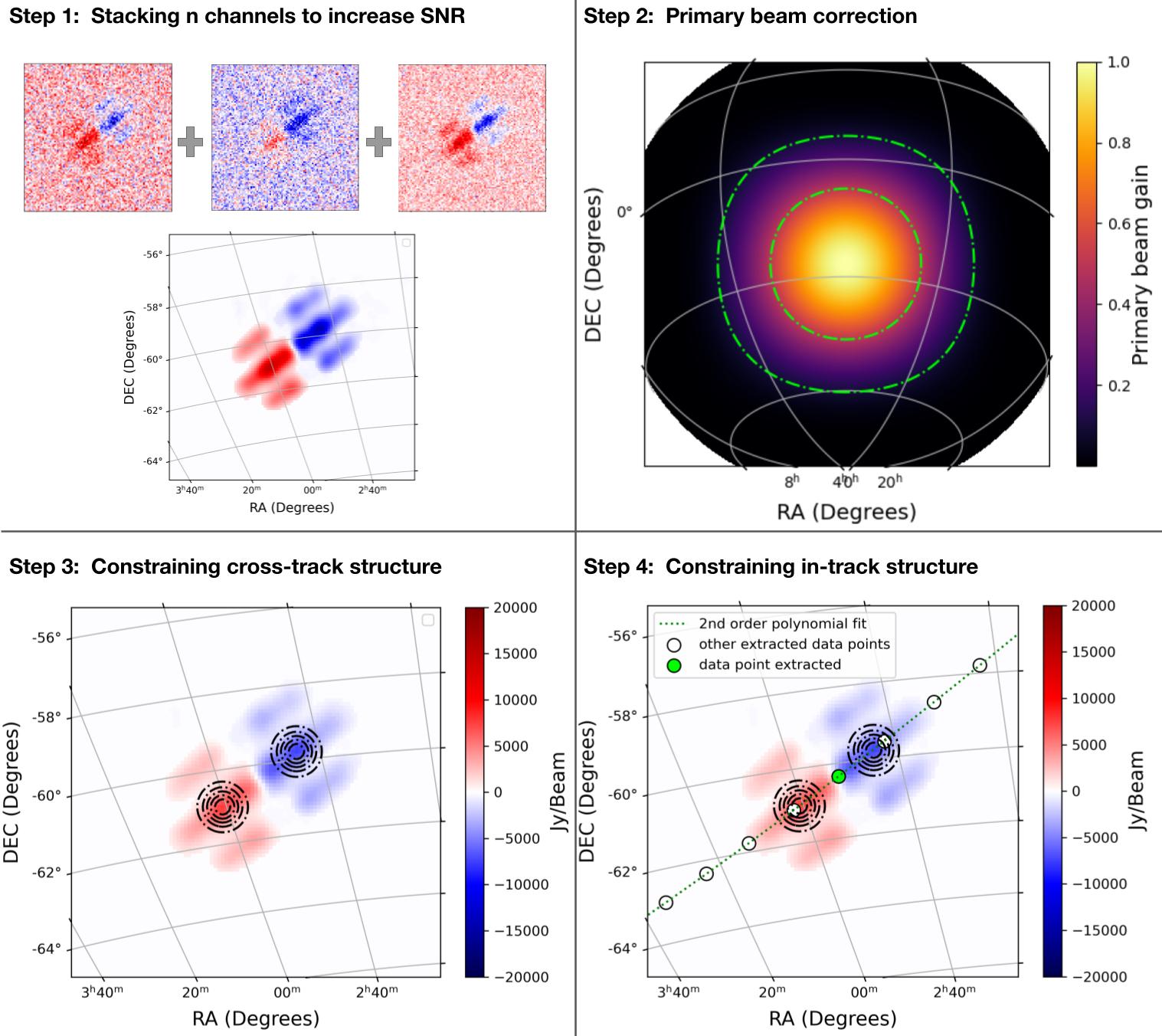}
\caption{The different panels of the figure show the steps involved in extracting angular position measurements described in Section \ref{sec:methods}. The top-left panel shows the different fine channels being stacked together to obtain a higher SNR streak signal. The top-right panel shows the primary beam of the MWA that is used to apply direction dependant gain corrections to the streak signal. The $50\%$ and $90\%$ of the primary beam response are shown using the green dashed lines.  In the bottom panels show the cross-track and in-track structure that is fit for in Section \ref{sec:crosstrack} and Section \ref{sec:intrack}. The Gaussian fits to the peak positive/negative pixels are shown as black dotted contours. The bottom-right figure also shows the extracted measurements (transition point) as a green circle along with other measurements from adjacent time-steps as white circles. }
\label{Fig2streak}
\end{center}
\end{figure*}

\begin{figure*}[h!]
\begin{center}
\includegraphics[width=\linewidth,keepaspectratio]{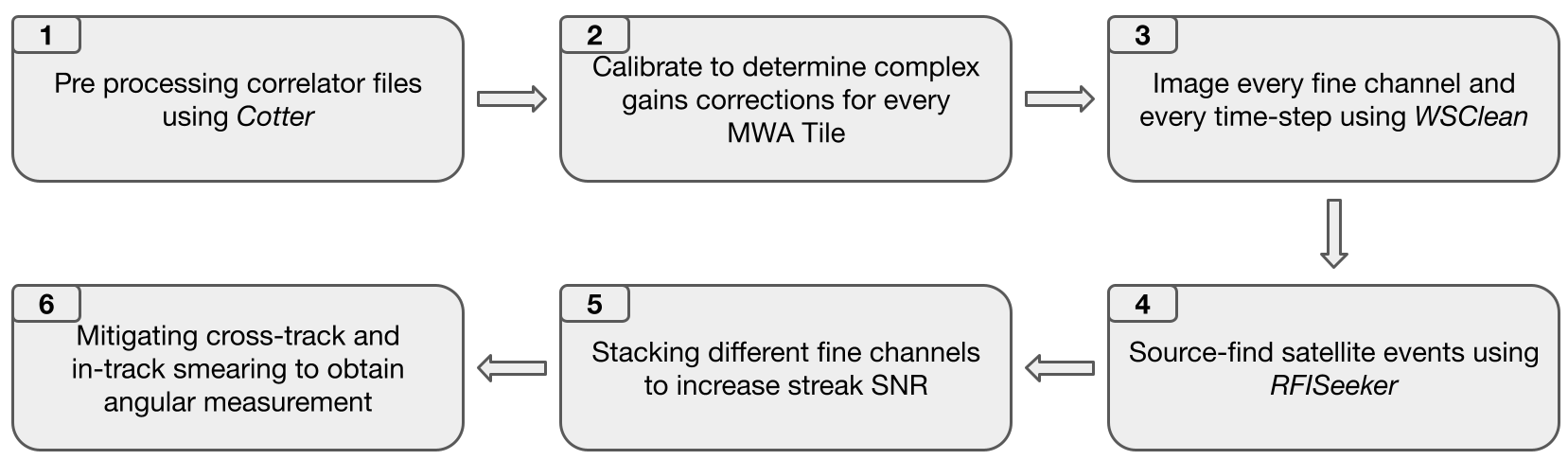}
\caption{Flow chart showing the different steps in the data reduction. In the first block, the correlator files are pre-processed into measurement set format. In block two, models of background radio galaxies is used to calibrate the phase and amplitude of the aperture array, followed by imaging every $40$\,kHz fine frequency channel at every $2$\,s intervals using {\tt WSClean} in block three. Blind detection of satellite events is then performed using source finding software, {\tt RFISeeker}, in block four and more information on steps 1-4 can be found in the previous work \citep{prabu_survey}. In block five, different channels with satellite signals are stacked to obtain a higher SNR signal in order to be able to do better astrometry in block six.  }
\label{Figflowchart}
\end{center}
\end{figure*}

In order to perform orbit estimation, multiple measurements of a satellite's angular position during the pass is obtained from the difference image data. The MWA's Phase 2 compact configuration \citep{2018PASA...35...33W} has a $\sim15$\, arcminute Full Width Half Max (FWHM) Point Spread Function (PSF) at FM frequencies, and thus the detected signals in difference images appear elongated in both in-track and cross-track directions, as can be seen in bottom panels of Figure \ref{Fig2streak}, due to PSF structure and the motion of the object during the observation period. These challenges are dealt with in three steps as described in Section \ref{sec:methods} in order to measure accurate time-stamped angular coordinates and then proceed to estimate the orbits in Section \ref{sec:orb} using these measurements. A high-level flow chart of the different steps in the data reduction is shown in Figure \ref{Figflowchart} and is further explained below. 

\subsection{SIGNAL RECONSTRUCTION}
\label{sec:reconstruction}
The previous work has observed satellites to reflect FM signals at multiple frequencies \citep{Tingay2013OnFeasibility,prabu_dev, prabu_survey}. Sometimes, for weak reflection events, different parts of the satellite track are visible at different frequencies. So, as step 1, the detections across different frequency channels are combined to obtain a higher signal to noise measurement, in order to reconstruct the satellite track more accurately via the aggregation of detections at different frequencies. The positive head and the negative tail of detected streaks (see top-left panel of Figure \ref{Fig2streak}) from different fine frequency channels are extracted and combined using the source finding software {\tt RFISeeker}\footnote{\url{https://github.com/StevePrabu/RFISeeker}. More information on the usage of the software can be found in \citep{prabu_survey}}, that was developed during the previous work \citep{prabu_survey}. The combined detection is then corrected for primary beam attenuation using a primary beam model \citep{sokolowski2017calibration} generated at the observed frequency.

\subsection{CONSTRAINING CROSS-TRACK STRUCTURE}
\label{sec:crosstrack}
 Since the curvature of the satellite's pass is not resolved by the MWA on the $2$\,s timescales of the difference images (as can be seen in Figure \ref{Fig2streak}), the cross-track elongation of the signal can be attributed to PSF structure alone. Hence, as step 2, the cross-track structure is characterised by fitting a model of the PSF to the detected streak. A Gaussian model of the PSF (statistical significance of the method is explained in \cite{condon1997errors}) is fit to the peak maximum (and minimum) pixel in the head (and tail) of the streak. The model fitting is performed using {\tt scipy.curvefit} \footnote{\url{https://docs.scipy.org/doc/scipy/reference/generated/scipy.optimize.curve_fit.html}} and the $1 \sigma$ uncertainties of the fit location are calculated using the square-root of the diagonal elements in the returned co-variance matrix. The average uncertainty in the Gaussian fit location was found to be under $5\%$ of the full width at half maximum (FWHM) of the PSF, approximately 0.75 arcminutes\footnote{the observations were phase calibrated using known positions of background galaxies, and in radio-astronomy, for a well-calibrated system, it is an accepted practice to use the error from Gaussian fitting as the astrometry error, and hence is also used to estimate the error in the angular measurements in this work. } (Note, this 1 arcminute accuracy is just an estimate, and would need to be proved through the analysis of residuals of measurements of objects with precise orbits generated through other means. This task is left as future work for the current analysis). Repeating the same step for the heads and the tails detected at multiple time-steps, constrains the orbital pass to an arc (with an average cross-track error less than $5\%$ of PSF FWHM).  Note that the location of the peak [maximum/minimum] pixel in the streak may depend on the altitude of the satellite as well as the FM illumination pattern of the apparent MWA sky. Hence, the time stamp information of these peak pixels cannot be determined without constructing a comprehensive model of the apparent FM illumination, and is beyond the scope of this work.
 
\subsection{CONSTRAINING IN-TRACK STRUCTURE}
\label{sec:intrack}

The in-track structure of the streak is mainly due to the motion of the satellite during the $2$\,s observation period. Since difference imaging is performed by subtracting the image at time-step $t-1$ from the image at time-step $t$ \citep{zhang2018limits,prabu_dev, prabu_survey}, the location of the satellite at the beginning of time-step $t$ corresponds to the point where the streak transitions from the negative tail to the positive head, and the corresponding time-stamp can be found using the header time information provided in the two images used\footnote{{\tt WSClean} \citep{offringa-wsclean-2014,offringa-wsclean-2017} is used to create MWA sky images and the software uses the mid UTC time of the integration as the \textit{OBS-DATE} parameter in the image header.}. Hence, as step 3, the final time-stamped angular measurements of the pass are obtained, by determining the point where the re-constructed streak signal transitions from head to tail along the constrained arc (as shown with green markers in the bottom-right panel of Figure \ref{Fig2streak}). Similarly doing so for every detected streak provides multiple timestamped angular measurements.  These values are used to perform preliminary orbit estimation in Section \ref{sec:orb}.

\begin{figure*}[]

\begin{center}
\includegraphics[width=\linewidth,keepaspectratio]{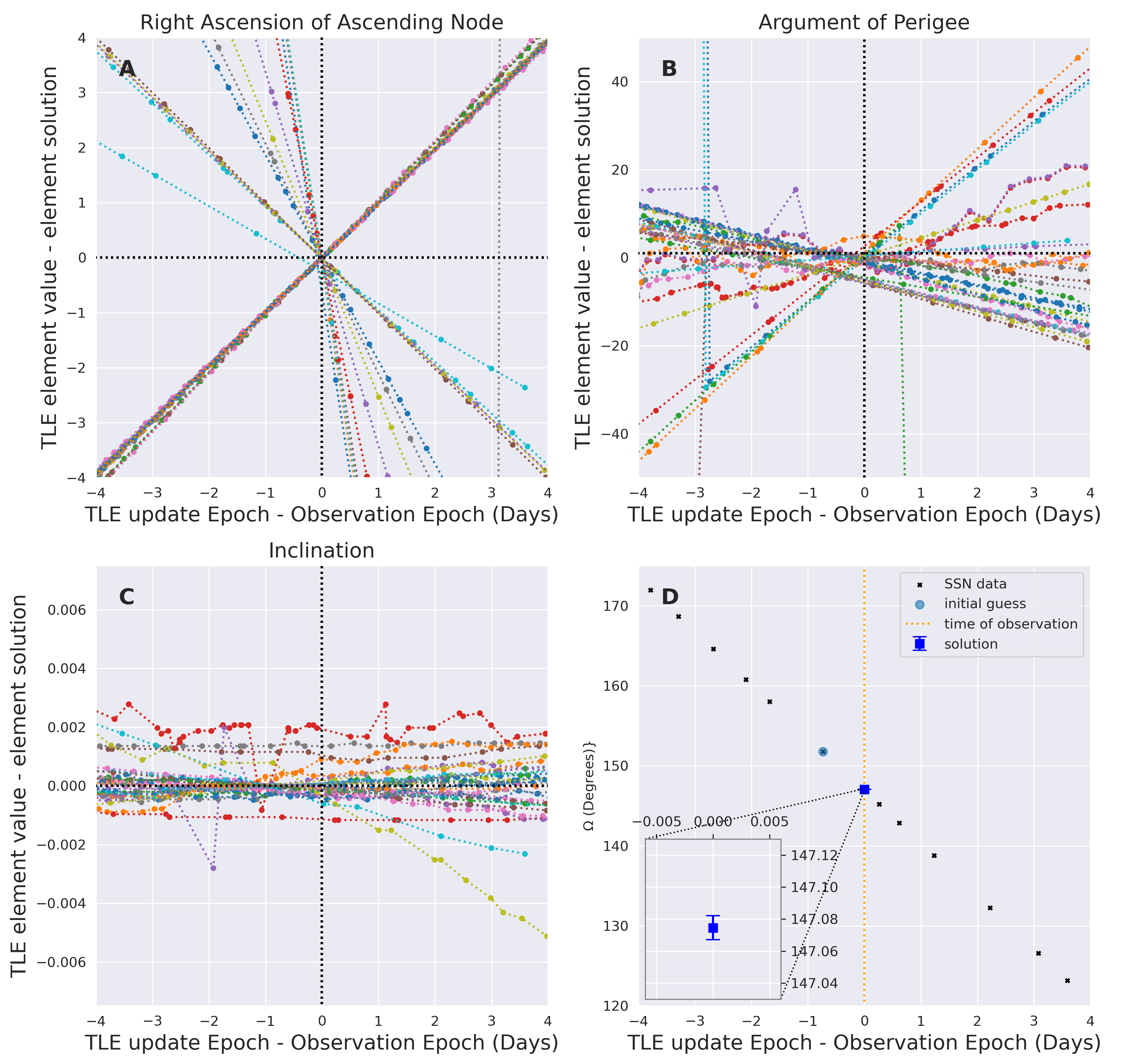}
\caption{In panels A, B and C, historic evolution of $\Omega$, $\omega$, and $i$ is shown for all the objects listed in Table \ref{tab1}, relative to the solutions. Panel D shows the historic values of $\Omega$ obtained near the epoch of observation (for the HST event detected in observation 1165762576). The initial guess used by the pipeline is shown as a blue circle marker, along with the determined value (with $2 \sigma$ uncertainties) at the epoch of the observation.}
\label{Fig5tle}
\end{center}
\end{figure*}

\begin{table*}[h!]
\caption{List of observations and target objects used for orbit determination in this work.}
\centering
{\footnotesize %
\begin{tabular*}{\textwidth}{@{} c\x c\x c\x c\x |c\x |c\x| c\x| c\x c\x  @{}}
\hline \hline
 Obs ID    &START UT      &  OBJ. NAME & RCS ($m^2$)   &  $\Omega$  & e &  $\omega$     & in-track (RMS km) \\ 
   DATE UT & STOP UT  & NORAD  &  $N_{\rm meas.}$ & M & $\eta$  &  i  & cross-track (RMS km) \\

\hline \hline
1157468632 & 15:03:35 & HST & 28.0799 & 61.675(2) & 0.000289(1) & 38.4(6) & 0.4 \\
2016-09-09 & 15:05:27 & 20580 & 12 & 232.5(6) & 15.0848526(4) & 28.4700 & 0.4 \\
\hline
1160497672 & 16:27:35 & ISS(ZARYA) & 399.0524 & 182.0035 & 0.000655(6) & 73.3056 & 0.8 \\
2016-10-14 & 16:29:27 & 25544 & 15 & 144.36648(3) & 15.54191(4) & 51.6421(2) & 2.3 \\
\hline
1160497672 & 16:27:35 & UKUBE-1 & 0.118 & 26.8606 & 0.0005614 & 118.1(3) & 0.4 \\
2016-10-14 & 16:29:27 & 40074 & 5 & 215.7(3) & 14.832312(3) & 98.3375 & 0.8 \\
\hline
1160497552 & 16:25:35 & ISS(ZARYA) & 399.0524 & 182.0104 & 0.0006740(1) & 73.29(3) & 0.4 \\
2016-10-14 & 16:27:27 & 25544 & 19 & 140.9332(3) & 15.54191(4) & 51.6421(2) & 0.2 \\
\hline
1160496232 & 16:03:35 & GPM-CORE & 8.104 & 34.749(1) & 0.0009966(1) & 273.28(2) & 1.9 \\
2016-10-14 & 16:05:27 & 39574 & 7 & 56.35(2) & 15.55243(6) & 65.01180(6) & 2.4 \\
\hline
1160495752 & 15:55:35 & ARIANE40+3R/B & 9.7046 & 202.8305 & 0.0016951(4) & 353.7765(5) & 0.5 \\
2016-10-14 & 15:57:27 & 23608 & 8 & 212.8078(7) & 14.959279(8) & 98.2094 & 0.9 \\
\hline
1160493592 & 15:19:35 & GAOFEN2 & 3.506 & 6.1414 & 0.0007181(1) & 200.8(2) & 1.2 \\
2016-10-14 & 15:21:27 & 40118 & 16 & 132.2(2) & 14.8058657(2) & 97.9507 & 1.0 \\
\hline
1160493472 & 15:17:35 & DUCHIFAT-1 & 0.037 & 195.3847 & 0.0014248(2) & 40.2530(3) & 0.4 \\
2016-10-14 & 15:19:27 & 40021 & 5 & 167.6723(9) & 14.895209(3) & 97.9214 & 0.7 \\
\hline
1160493472 & 15:17:35 & SPOT4 & 6.193 & 5.60(1) & 0.0012212 & 313.5(4) & 9.9 \\
2016-10-14 & 15:19:27 & 25260 & 23 & 20.6(4) & 14.53434552 & 98.3715 & 14.3 \\
\hline
1160487832 & 13:43:35 & OKEAN-4 & 7.1433 & 349.2752 & 0.0020528(1) & 160.9(6) & 1.8 \\
2016-10-14 & 13:45:27 & 23317 & 8 & 171.8(6) & 14.872731(1) & 82.53949(7) & 0.1 \\
\hline
1160485792 & 13:09:35 & SHIJIAN-16(SJ-16) & 8.2723 & 148.82350(2) & 0.0018570(6) & 102.9(4) & 0.8 \\
2016-10-14 & 13:11:27 & 39358 & 9 & 104.5(4) & 14.8528152(1) & 74.97430(4) & 0.1 \\
\hline
1165766176 & 15:55:59 & KANOPUS-V1 & 1.9 & 263.8922 & 0.0001989 & 70.2475(6) & 0.1 \\
2016-12-14 & 15:57:51 & 38707 & 15 & 136.8321(6) & 15.2001829(1) & 97.4517 & 1.2 \\
\hline
1165762576 & 14:55:59 & HST & 28.0799 & 147.0746(3) & 0.00027370(2) & 351.1(8) & 0.7 \\
2016-12-14 & 14:57:51 & 20580 & 10 & 286.0(8) & 15.08620(3) & 28.46880(7) & 0.1 \\
\hline
1165761136 & 14:31:59 & ZIYUAN3-1(ZY3-1) & 5.315 & 55.5670 & 0.00010559(8) & 74.7(2) & 1.1 \\
2016-12-14 & 14:33:51 & 38046 & 14 & 257.59(2) & 15.213742(3) & 97.3105 & 1.0 \\
\hline
1165760776 & 14:25:59 & COSMOS1328 & 8.2828 & 60.1338(1) & 0.00111030(2) & 167.3(1) & 0.6 \\
2016-12-14 & 14:27:51 & 12987 & 11 & 165.7(1) & 15.071518(4) & 82.516(0) & 0.0\\
\hline
1165755976 & 13:05:59 & COSMOS1544 & 8.2989 & 213.2614 & 0.0014169(0) & 341.81638(2) & 1.8 \\
2016-12-14 & 13:07:51 & 14819 & 16 & 226.1142(5) & 15.25249645(1) & 82.48479(1) & 0.8 \\
\hline
1165752856 & 12:13:59 & SPOT1 & 7.279 & 21.129(2) & 0.0141920(5) & 257.8539(1) & 0.7 \\
2016-12-14 & 12:15:51 & 16613 & 7 & 73.5564 & 14.6484554(1) & 98.7430(1) & 1.2 \\
\hline
1165771216 & 17:19:59 & GOSAT(IBUKI) & 4.6494 & 97.0669 & 0.0001526 & 91.8756 & 13.0 \\
2016-12-14 & 17:21:51 & 33492 & 3 & 241.45160(2) & 14.67526130(1) & 98.1110 & 11.7 \\
\hline
1165753936 & 12:31:59 & COSMOS1766 & 8.2879 & 204.2835 & 0.001571(1) & 31.6(1) & 0.7 \\
2016-12-14 & 12:33:51 & 16881 & 14 & 175.5(1) & 15.09860(8) & 82.50508(4) & 0.2 \\
\hline
1165753936 & 12:31:59 & ATLAS2CENTAURR/B & 14.8664 & 122.91(1) & 0.642931(4) & 248.20(2) & 0.3 \\
2016-12-14 & 12:33:51 & 23968 & 20 & 3.079(3) & 3.365053(5) & 26.72361(1) & 0.2 \\
\hline
1165773496 & 17:57:59 & YAOGAN24 & 4.2274 & 105.7079 & 0.0016264 & 309.38(3) & 0.8 \\
2016-12-14 & 17:59:51 & 40310 & 5 & 24.7(3) & 14.7706137(2) & 97.98487(4) & 1.6 \\
\hline
1165771096 & 17:17:59 & FGRST(GLAST) & 4.9326 & 179.10(5) & 0.0012252(2) & 308.6(1) & 0.6 \\
2016-12-14 & 17:19:51 & 33053 & 7 & 331.0(1) & 15.107828(2) & 25.58160(1) & 0.2 \\

\hline \hline

\end{tabular*}\label{tab1}
\tabnote{continued on next page...}
}%

\end{table*}

\addtocounter{table}{-1}
\begin{table*}[h!]
\caption{...continued from previous page.}
\centering
{\footnotesize %
\begin{tabular*}{\textwidth}{@{} c\x c\x c\x c\x |c\x |c\x| c\x| c\x c\x  @{}}
\hline \hline
 Obs ID    &START UT      &  OBJ. NAME & RCS ($m^2$)   &  $\Omega$  & e &  $\omega$     & in-track (RMS km) \\ 
   DATE UT & STOP UT  & NORAD  &  $N_{\rm meas.}$ & M & $\eta$  &  i  & cross-track (RMS km) \\ 
\hline \hline

1165764136 & 15:21:59 & IRS-P4(OCEANSAT) & 3.4542 & 67.0832 & 0.00043390(5) & 114.3260(9) & 1.2 \\
2016-12-14 & 15:23:51 & 25758 & 6 & 219.0281(1) & 14.523170(7) & 98.17660(1) & 1.3 \\
\hline
1165761376 & 14:35:59 & H-2AR/B & 27.4086 & 184.5777 & 0.00182159(2) & 80(1) & 0.5 \\
2016-12-14 & 14:37:51 & 41341 & 8 & 157.7(1) & 15.040126(7) & 30.6257(3) & 0.2 \\
\hline
1165761256 & 14:33:59 & ZIYUAN3-1(ZY3-1) & 5.315 & 55.5683 & 0.00010559(1) & 74.49(2) & 0.8 \\
2016-12-14 & 14:35:51 & 38046 & 5 & 260.44(2) & 15.213743(5) & 97.31052(5) & 0.7 \\
\hline
1165758616 & 13:49:59 & METOP-A & 11.2479 & 46.0404 & 0.00013536(4) & 48.530(7) & 0.6 \\
2016-12-14 & 13:51:51 & 29499 & 5 & 284.656(7) & 14.21485052(3) & 98.7065 & 1.1 \\
\hline
1165757056 & 13:23:59 & ENVISAT & 18.597 & 36.2852 & 0.0001173 & 83.980(8) & 1.0 \\
2016-12-14 & 13:25:51 & 27386 & 11 & 249.480(8) & 14.3788361(8) & 98.25991(3) & 1.4 \\
\hline
1160491192 & 14:39:35 & COSMOS2082 & 10.7604 & 8.5789(6) & 0.00136930(2) & 35.1(8) & 0.3 \\
2016-10-14 & 14:41:27 & 20624 & 3 & 296.4(8) & 14.14209(2) & 71.04320(2) & 0.1 \\
\hline
1160484112 & 12:41:35 & ATLAS3BCENTAURR/B & 11.93 & 43.4793 & 0.3205425(1) & 303.20(1) & 1.5 \\
2016-10-14 & 12:43:27 & 28118 & 4 & 351.242(9) & 9.06535(6) & 28.1480(1) & 0.8 \\
\hline
1157474632 & 16:43:35 & HST & 28.0799 & 60.629(1) & 0.00029159(1) & 39(1) & 1.3 \\
2016-09-09 & 16:45:27 & 20580 & 18 & 253.2(1) & 15.084844(6) & 28.469(2) &  0.1\\
\hline
1157472832 & 16:13:35 & RESURSP3 & 7.66 & 346.0609 & 0.00042382(9) & 20.8(3) & 0.7 \\
2016-09-09 & 16:15:27 & 41386 & 3 & 311.8(3) & 15.3223449(3) & 97.2829(4) & 0.9 \\
\hline
1157407072 & 21:57:35 & RADARSAT-2 & 8.381 & 258.6263 & 0.0001236(7) & 95.28(2) & 0.6 \\
2016-09-08 & 21:59:27 & 32382 & 4 & 110.8(2) & 14.29982302 & 98.57728(2) & 1.5 \\

\hline \hline
\end{tabular*}\label{tab1}
}%
\tabnote{The observation IDs\footnote{The observations used are publicly available and can be obtained from \url{https://asvo.mwatelescope.org/}} for all the satellite passes used in this work. The table also lists the corresponding UTC time for the observations, along with the NORAD ID for the satellite detected in the observation, and its Radar Cross-Section (RCS)\footnote{\url{https://celestrak.com/pub/satcat.txt}}. $N_{\rm meas.}$ is the number of detections of the object during the observation period.  The determined orbital elements (along with the 1$\sigma$ error on the last digit within brackets) are reported for each of these objects and the RMS in-track and cross-track residuals (in arc-minutes) after the fit. No. measurements is the number of angular position measurements obtained for each event.}
\end{table*}

\begin{figure*}[h!]
\begin{center}
\includegraphics[width=\linewidth,keepaspectratio]{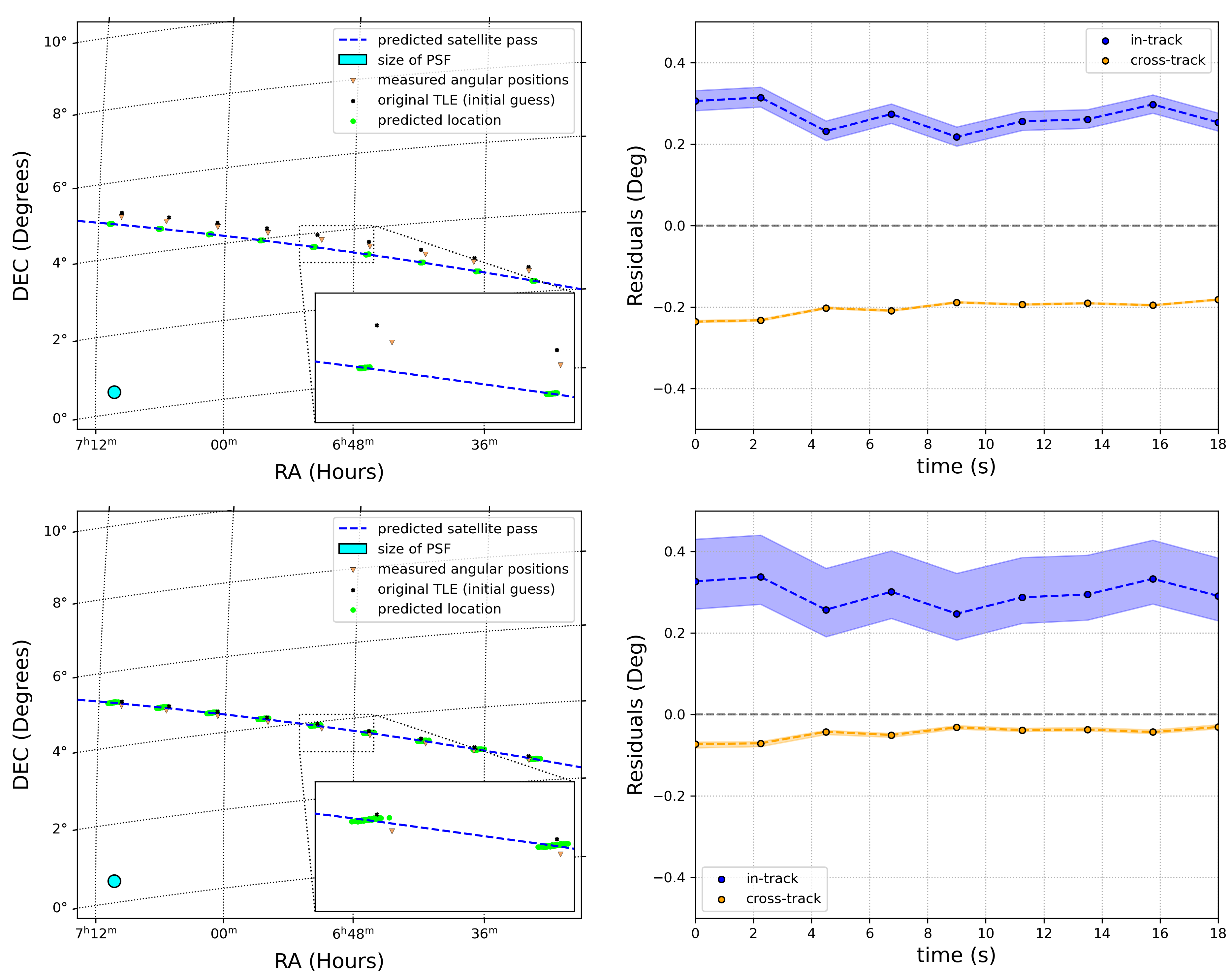}
\caption{The measured HST pass using observation $1165768696$ along with the predictions derived using measurements from a single prior orbit (top-left and top-right panels) and derived from measurements from two prior orbits (bottom-left and bottom-right panels). The predicted trajectory of the pass using the orbital element solutions are shown as a blue dashed line. The measured angular positions of the satellite are shown as orange triangle markers and the predicted positions through MC sampling of the errors associated with the orbital elements are shown as green circle markers. The residuals of the fit projected in in-track and cross-track directions are shown in the right panels. The approximate size of the zenith pointed PSF of the observation is shown as a cyan ellipse in the bottom left of top-left and bottom-left panels. }
\label{Fig7satelliteReacquisition}
\end{center}
\end{figure*}

\section{RESULTS}
\label{sec:results}

\subsection{DETERMINING THE ORBITAL ELEMENTS}
\label{sec:orb}
Different types of SDA sensors exist: radars, providing angular measurements and slant range and/or Doppler measurements; on-ground lasers, providing line-of-sight range measurements, and telescopes, providing angular measurements only. Most modern SDA radars can perform orbit determination using the derived position and velocity vectors. For SDA telescope detections, angles-only orbit determination can also be performed using methods such as the Gauss, Laplace \citep{curtis2013orbital}, and Gooding methods \citep{henderson2010modifications}. However these methods do not work under all scenarios for fast moving LEO objects (as also described by \citep{e1e6a6efa7824061a09826100bf5489a} and \cite{wijnen2020using}), and are more used with success for heliocentric objects with smaller apparent angular speed such as asteroids, comets, and Kuiper belt objects \citep{10.1007/978-1-4020-5325-2_19}. The non-coherent passive radar techniques used in this work are confined to angles-only measurements. Upon testing angles-only initial orbit determination methods, poor solutions for $e$, $M$, and $\omega$ was often obtained, and for this reason, a least squares fitting starting from a transit prediction is exploited. 

In order to use the measurements for orbit estimation, the orbital elements are constrained by performing a least-squares fitting the predicted trajectory with the angular measurements using a python implementation\footnote{\url{https://rhodesmill.org/skyfield/}} of a Simplified General Perturbations (SGP4) propagator. The cost function used is the sum of squared residuals between the observed and the predicted angular positions in the sky.  The publicly available Two-Line Element (TLE) descriptions of the orbital elements were obtained for the objects selected in this work\footnote{using the API query feature of \url{space-track.org}} and the TLE closest in time prior to the observations is used as the initial guess for the object's orbital elements is used (note, all TLEs of the previous approx. two weeks can also be used as an initial guess, as they were all found to converge on similar solutions as demonstrated in Section \ref{sec:validation}). Orbital element $M$ is a derived measure of the time elapsed since the object has crossed the perigee, and hence is a parameter that is dependant on the location of the observer and the location of the line of periapsis. Hence, for $M$ the initial guess was calculated using Equation \ref{E2}.

\begin{equation}
    M_{1} = (t_{1} - t_{0})\times \eta_{0} \times 360 + M_{0}\\
    \label{E2}
\end{equation}

where $M_{0}$ is the mean anomaly from a previous TLE update, $(t_{1} - t_{0})$ is the time elapsed since the observation time from the previous TLE update\footnote{The UTC time of TLE update can be determined using the "EPOCH" parameter that is returned when using the {\tt tle\_query} class from the {\tt spacetracktool} \url{https://pypi.org/project/spacetracktool/} python module}, and $\eta_{0}$ is the mean motion from the previous observation. Since $\eta$ is the average number of revolutions per day, and because $M$ increases linearly with time (irrespective of the orbit being circular or elliptical), the initial guess for $M$ at time $t$ can be determined using Equation \ref{E2}.

The orbital parameters were fit in two steps.  First, allowing only the elements with large expected variations (elements that undergo large changes with time) are allowed to vary (i.e, $\Omega$, $M$, and $\omega$) to find the global minimum using {\tt scipy.basinhopping}\footnote{\url{https://docs.scipy.org/doc/scipy/reference/generated/scipy.optimize.basinhopping.html}}. Second, fine-tuned adjustments is performed for all six elements using {\tt scipy.curve\_fit}\footnote{\url{https://docs.scipy.org/doc/scipy/reference/generated/scipy.optimize.curve\_fit.html}}.

To encourage convergence of a multi-dimensional ($6$ orbital parameters) least-squares fit, boundary conditions are set for the six elements in both the fitting steps. The boundary condition for an orbital element is determined by inspecting the maximum variation in the orbital element value over the course of the past 60 days. This method seemed to provide good limits not just for passive objects, whose orbits primarily change due to atmospheric drag and J2 effects alone, but also for active objects with manoeuvring capabilities. The orbit estimates obtained for each of the $32$ satellite passes are given in Table \ref{tab1}. The orbital elements up to the precision required to generate TLEs are reported. 

From Section \ref{sec:crosstrack}, the typical positional uncertainty in the individual angular position measurements is known to be less than an arcminute. Hence, the $1\sigma$ uncertainties associated with the orbital elements is calculated by performing a Monte-Carlo (MC) \citep{metropolis1949monte} re-sampling from a 2D normal distribution of angular positions (with standard deviation equal to the error in the angular position measurement).  The estimated uncertainties are listed in Table \ref{tab1}. For some of the objects in Table \ref{tab1}, more position measurements than for other objects were obtained (due to being detected in more time-steps), and hence were able to better constrain the orbital elements with smaller uncertainties. Also, due to using 2D projected data (RA-DEC) of the 3D motion of the satellite, the uncertainties were often found to be highly coupled (for example, $\Delta M$ and $\Delta \omega$ reported in Table \ref{tab1} were found to be almost always equal in value). The trace of $M$ and $\omega$ during the MC sampling almost always had a correlation coefficient of approx. $-1$. This is also due to the orbits being almost circular, and due to perigee in a circular orbit being not defined. The residuals ($fit - measured$) for every orbit determination are projected into in-track and cross-track directions and the RMS of the residuals in these two directions are listed in Table \ref{tab1} \footnote{A demo pipeline of the scripts used to perform angular measurement extraction and orbit determination can be found in \url{https://github.com/PhD-Misc/MWA-OrbitDetermination}.}.

\subsection{VALIDATION OF RESULTS}
\label{sec:validation}
The orbital elements obtained from the method were verified against the TLE values available via the SSN near the epoch of observation. The values were in close agreement with the orbital elements extracted from the TLEs.  For all the events detected in Table \ref{tab1}, the fractional historic evolution of their orbital elements ($\Omega$, $\omega$ and $i$) is shown in panels A, B and C of Figure \ref{Fig5tle}. Most of the parameter trends go through the origin (they don't have to as the influence of drag on a tumbling LEO object with varying drag ram area is very complex) and the solutions are in good agreement with the public TLE values from near the epoch of observation. 

An example for one of the orbital elements ($\Omega$ for HST pass during the observation $1165762576$) is shown in panel D of Figure \ref{Fig5tle}. Figure \ref{Fig5tle}.D also shows the initial guess used by the pipeline, and the $\Omega$ solution with $2 \sigma$ uncertainty. Although an initial guess is used to seed the orbit determination, the dependence of the converged solution is tested and turns out to be independent of the initial guess (more information in Section \ref{sec:convergence}). Nevertheless, using an initial guess helps converge to the solution faster and helps save computation costs. 

\subsection{HST re-acquisition}
\label{sec:reacquisition}
During one of the four observation periods used during the LEO blind survey using the MWA \citep{prabu_survey}, the HST was observed for 3 passes (more information in Table \ref{tab2}) on one of the nights.  These observations are used to test for re-acquisition of the satellite based on the estimated orbital elements.  

\begin{table*}[h!]
    \caption{Information on the HST observations described in Section \ref{sec:reacquisition}}
    \centering
    \begin{tabular}{@{}ccccc@{}}
    Obs ID & Start  & Stop & time-steps   & additional comment\\
           &  UTC   & UTC  & detected &  \\
    \hline \hline
        1165756576 &2016-12-14 & 2016-12-14& 3 & Used for 2-pass orbit determination (Section \ref{sec:2pass}).  \\
        &13:15:59.0 &13:17:51.0 & &  \\\hline 
        1165762576 &2016-12-14 &2016-12-14 & 10 & Used for 1-pass orbit determination (Section \ref{sec:1pass}). \\
        &14:55:59.0 & 14:57:51.0& & Used for 2-pass orbit determination (Section \ref{sec:2pass}) \\ \hline
        1165768696 & 2016-12-14& 2016-12-14&13 & Used to test for target re-acquisition\\
        &16:37:59.0 &16:39:51.0 & &  using the determined orbital elements\\\hline 
       
    \end{tabular}
    \label{tab2}
\end{table*}

\subsubsection{One pass re-acquisition}
\label{sec:1pass}
Using the angular position measurements obtained for the HST pass observed during the observation $1165762576$ (part of Table \ref{tab1}), orbit determination is performed and the obtained solutions are used to predict the subsequent HST pass visible to the MWA (observation $1165768696$). The two observations are one orbit apart (approximately $100$\, minutes). Panel A of Figure \ref{Fig7satelliteReacquisition} shows the predicted trajectory using the determined orbital elements as the blue dashed line and the predicted positions (using MC sampling of the orbital elements) as green markers. The predicted trajectory and the measured trajectory (inverted red triangles) match to within $0.2^{\circ}$ (1.9 km at the HST altitude) in the cross-track direction and $0.3^{\circ}$ (2.8 km at the HST altitude) in the in-track direction. The in-track and cross-track offsets between the predicted angular position and the measured positions of each of the MC walkers are calculated and plotted in Panel B of Figure \ref{Fig7satelliteReacquisition}.


\begin{figure*}[]

\begin{center}
\includegraphics[width=\linewidth,keepaspectratio]{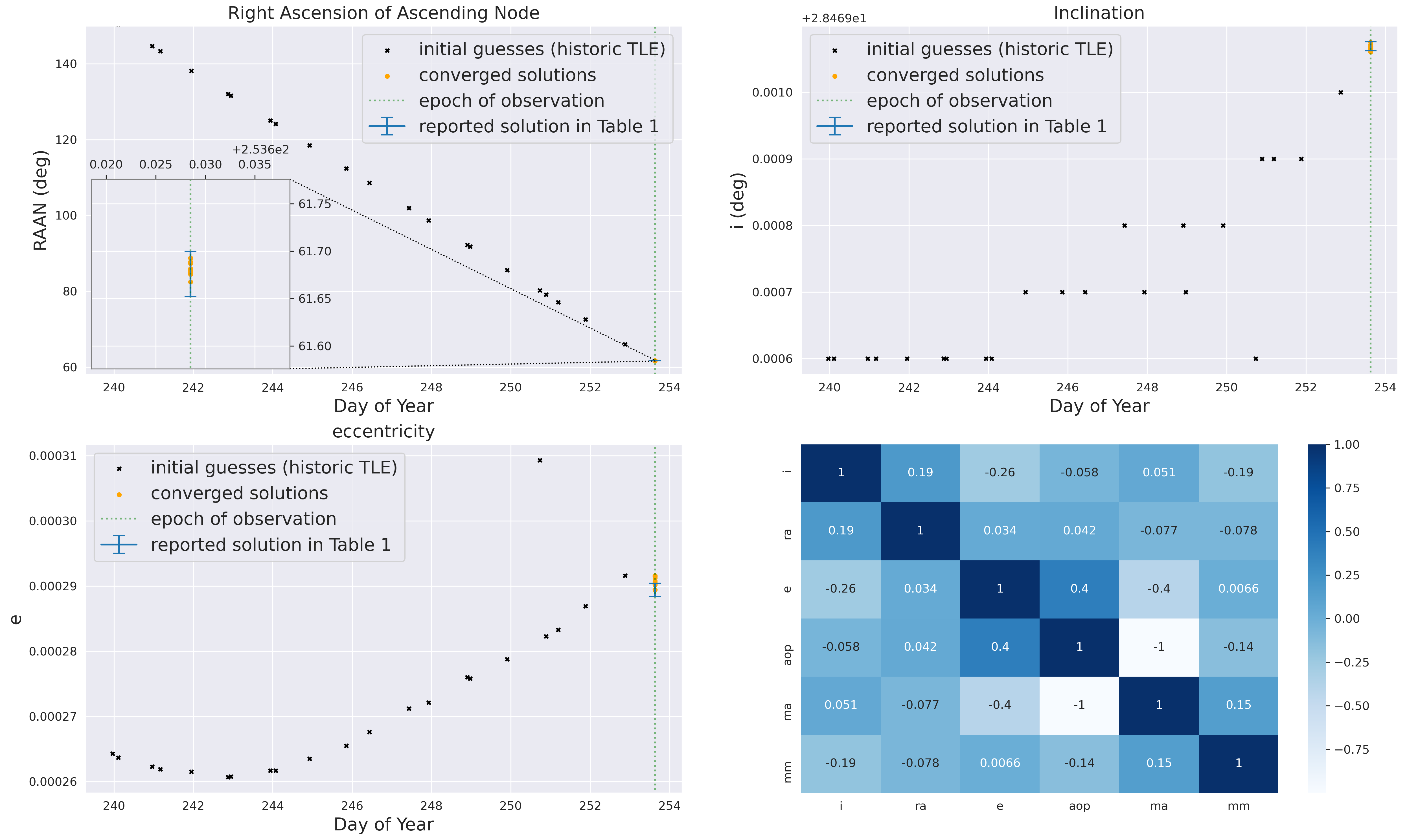}
\caption{Panels A, B, and C show different initial guesses (for $\Omega$, $i$, and $e$ respectively) provided to the orbit determination pipeline using black scatter points. The converged solution for all the provided initial guesses are shown using orange scatter points. The solution determined in Table \ref{tab1} using closest epoch TLE as the initial guess is shown using blue error bars. The bottom-right panel shows the covariance between the different orbital elements.}
\label{Fig6conv}
\end{center}
\end{figure*}

\begin{figure*}[h!]
\begin{center}
\includegraphics[width=\linewidth,keepaspectratio]{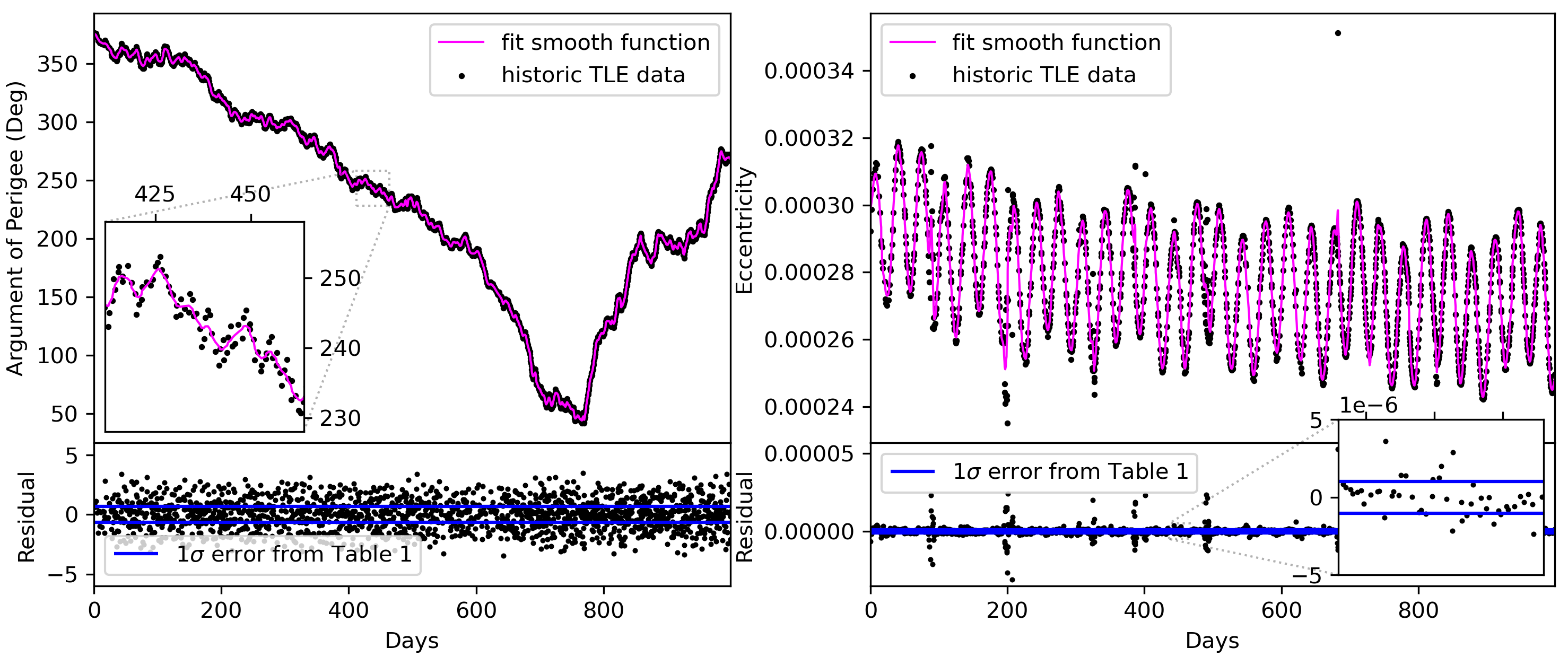}
\caption{Figure shows the historic evolution of $\omega$ and $e$ for HST over a 1000 days. The residual panel of each figure shows the estimated random error in the reported TLE updates along with the reported errors (in Table \ref{tab1}) for $\omega$ and $e$. Note that the order of magnitude of the scatter in $e$ that is of interest in this work is shown in the insert panel.}
\label{Fig8errorAnalysis}
\end{center}
\end{figure*}

\subsubsection{Two pass re-acquisition}
\label{sec:2pass}
The orbit determination method is then tested by using the angular position measurements of two HST passes (using observation $1165762576$ used for one pass re-acquisition and $1165756576$ observation that observed HST $100$ minutes prior to the former observation) to perform a joint orbit determination. The obtained orbital elements from the two-pass orbit determination were tested by attempting to re-acquire HST in a future pass and the prediction trajectory is shown in Panel C of Figure \ref{Fig7satelliteReacquisition}. The in-track and cross-track offset for the 2-pass satellite re-acquisition is shown in Panel D of Figure \ref{Fig7satelliteReacquisition}.

The observed reduction in cross-track offsets to less than 0.1$^{\circ}$ ($1$ km at the HST altitude) between the two methods is analogous to the improvement observed by optical studies \citep{bennett2015analysis} that compared prediction accuracy when using a short arc detection and a long arc detection to perform orbit determination. Although the MWA is a wide FOV sensor that is capable of detecting satellite passes that span more than $10\,$s of degrees, the curvature of the pass is not highly constrained due to arc-minute angular resolutions. Hence, using position measurements from multiple passes helps determine the orbital elements more accurately. The two-pass orbit determination also resulted in very small uncertainties for the $\eta$ orbital element, as multi-pass observations help tightly constrain the orbital period of the object. However, the in-track residual errors for two-pass re-acquisition seem to be larger than the single-pass case. This is attributed to combining angular position measurements from different parts of the MWA FOV during the two passes. The MWA baselines (an antenna pair) see a direction-dependant differential phase offset (resulting in small source position offsets) due to viewing the source through different ionosphere layers. As it is computationally expensive to perform Direction dependant calibration and is not performed here.

\subsubsection{ORBIT DETERMINATION SOLUTION CONVERGENCE}
\label{sec:convergence}

As described in Section \ref{sec:orb}, the orbit determination pipeline used in this work uses the closest epoch SSN published TLE for the target as the initial guess. In this section the dependence of the converged solution is tested and confirms that the pipeline is in-fact updating the TLE to the epoch of the observation and not re-confirming the initial guess provided. The HST pass in observation 1157468632 is used as the test example for this analysis. Rather than feeding the pipeline with the closest epoch TLE as the initial guess (as done in Section \ref{sec:orb}), the TLE published in the previous 14 days as the initial guess is proved and the converged solution is analyzed.

In Figure \ref{Fig6conv} the different initial guesses provided to the pipeline are shown using black markers and it can be noted that the orbital elements can vary substantially even within a few days. The converged solution for all the provided initial guesses is shown using orange scatter points along with the solution reported in Table \ref{tab1}. The converged solution is found to be independent of the initial guess and using the closest epoch TLE as the initial guess only helps the solution converge faster.

\section{DISCUSSION AND FUTURE WORK}
\label{sec:discussion}
\subsection{SDA Catalog Maintenance using the MWA and Future Improvements}

\subsubsection{SDA Sensitivity Dependence on Array Configuration and Hardware Parameters}
In this work, the compact configuration of the MWA Phase 2 array is used, which contains only short baselines.  The angular resolution of the  position measurements could be improved by using the Phase 2 extended configuration of the MWA (longest baseline spans approx. $5$\,km). However, upon testing SDA observations using the extended configuration, two limiting factors that reduced the sensitivity towards LEO objects are identified as the near-field effects and visibility fringe washing.  These effects are documented below. 

Interferometer theory assumes that the received wave-front from the source of interest is planar, i.e, the emitting/reflecting source is far away. However, due to the near-field nature of SDA targets, the curvature of the received wavefront can be observed using the longer baselines of the MWA extended configuration. Using the far-field equation obtained from \cite{zhang2018limits} ($d=2D^{2}/\lambda$, where $d$ is the distance of the object, $D$ is the baseline length, and $\lambda$ is the wavelength of the observation), objects at an altitude of $400$\, km (e.g ISS) were found to appear in the near-field for baselines longer than approx. $1160$\,m. Thus, for extended configuration SDA observations, due to the signal not being correlated coherently, a de-focusing effect was obtained that smears the signal over a large patch of the sky. 

When going from the short baselines of the compact configuration used in the current work towards longer baselines of the extended configuration, even before the near-field limitation comes into effect, visibility fringe washing is expected to become significant. The phase of the measured visibility contains information about the source position (assuming a single source) during the correlator time-averaging, and the measured phase changes with time as the source moves (usually due to sky rotation). The time-averaging interval of the correlator is often optimised to avoid spatial smearing of the data due to sky rotation and for existing MWA hardware this is limited to $0.5$\,s. However, in the SDA observations, LEO objects often have very high angular speeds (e.g ISS moves at approx. $1^{\circ}/s$ near the zenith), and hence due to rapid changes in the phase of the visibility, the $0.5$\,s time-averaging results in visibility fringe washing. A simplified form of fringe frequency (time rate of change in visibility phase) for an East-West baseline is given below (adapted from \cite{marr2015fundamentals})

\begin{equation}
    F_{freq} = 2\pi \omega \frac{b}{\lambda} cos(\omega \Delta t)\\
    \label{E3}
\end{equation}

where $F_{freq}$ is the fringe frequency, $\omega$ is the angular speed of the source, $b$ the projected baseline length, $\lambda$ the wavelength of the observing frequency and $\Delta t$ is the correlator integration time. Using Equation \ref{E3} it can be seen that the phase of the visibility changes by more than $\pi$ for an object like the ISS for baselines longer than $195$\,m, with the existing MWA hardware ($\Delta t=0.5$\,s). With the current ongoing upgrade to MWA Phase 3, the correlator time-averaging can reduce to $0.1$\,s and this should help increase the number of long baselines that can use for SDA observations.

\subsubsection{The MWA SDA Capability in the Global Context of SDA Sensors}

In this section, the MWA passive radar system is placed in the context of existing global SDA sensors. This is done by discussing three aspects of the system: the accuracy of position measurements; goodness of orbit determination and associated errors; and the value added to existing global SDA networks. 

The SDA system is able to perform angular position measurements with average uncertainties of $1.1$ arc-minutes, which are sufficiently small to support meaningful coordination with much smaller field-of-view optical sensors (such as the ZIMLAT telescope used in \cite{cordelli2019use} [$7'\times7'$ FOV] and the OWL-Net SDA sensor used in  \cite{choi2018optical} [$1.1^{\circ}\times1.1^{\circ}$ FOV]). Unaided laser ranging systems require a few arc-second accuracies \citep{bennett2015analysis}.  With the current MWA SDA system, coordinating with laser devices could prove to be a challenging task. The $1.1$ arc-minute angular error translates to under $0.5$\,km positional error at a distance of $1000$\,km. These typical uncertainties are comparable to the notional $\sim1$\, km positional errors \citep{vallado2013improved} of TLEs during the epoch of measurement.

In previous sections accuracy of the orbit determination method is validated by testing for re-acquirability of the target object, as well as by verifying them against the TLE values released by SSN near the epoch of observation. The uncertainties associated with these SSN TLE updates are not publicly available, and hence the reported orbital element errors (Table \ref{tab1}) are placed in the global context using the HST observation (observation $1165762576$ in Table \ref{tab1}) as an example. The historic TLE values for HST for $1000$ days are obtained and the orbital elements are plotted as a function of time ($\omega$ and $e$ shown in Figure \ref{Fig8errorAnalysis}). The SSN reported TLE values change with every entry due to the random errors associated with the sensor system that reports it, as well as due to the systematic drift in the orbital element value with time due to the impact of drag, and orbital maneuvers performed by the satellite. Since only the errors associated with the global SDA systems is of interest, the scatter in the data is isolated from the drift by fitting a smooth function to the data \footnote{performed using {\tt scipy.ndimage.gaussian\_filter1d}. The filter parameters were optimised by  trial and error until the residual in the fit showed no structure.}. The difference between the model and data can be used as an estimate of the errors in the reported values. In Figure \ref{Fig8errorAnalysis} the reported errors for $\omega$ and $e$ are shown with reference to the estimated errors in the TLE updates, and they are very similar in value. Note that in the residual panel for $e$ in Figure \ref{Fig8errorAnalysis}, the jumps/spikes in the residuals are probably due to maneuvers performed by the HST, and the random errors that are of interest in this work are shown in the insert residual panel.

With the growing density of objects in LEO, multi-target tracking capability is preferred for SDA sensors \citep{jones2015challenges}. The MWA SDA system is not just able to perform simultaneous detections, but also perform SDA observations at all times and is not constrained to observation windows like optical sensors that are confined to twilight. The large FOV of the MWA allows detections to be performed over a span $10$\,s of degrees across the sky. The difference in the number of detections obtained by the system when compared to active SDA radars is compensated by the lower operation cost involved in the system due to the exploitation of non-cooperative terrestrial FM transmitters. Most of the non-coherent SDA work done thus far (including \cite{prabu_dev} and \cite{prabu_survey}), was performed using archived data, meaning the primary science case for the observation was not to perform SDA. Observations designed specifically for SDA, using an optimised and dedicated facility based on the MWA technologies and the techniques, would prove to be even more productive.

\subsection{Importance of FM Reflecting LEO Catalog for Astronomy Observations}
The MRO is not just the home to the MWA, but also to other interferometers that operate at FM frequencies, such as the Engineering Development Array (EDA) \citep{wayth2017engineering} and the future low-frequency component of the SKA. Maintaining a catalog of LEO objects that are known to reflect FM transmitters can help understand FM band observations performed using these low-frequency radio interferometers. Not all optically "small" objects are "small" objects for low-frequency observations. The RCS of a "small" object can drastically increase by the presence of a broken antenna or a wire, and thus maintaining a catalog of radio-bright LEO objects is not just useful for SDA purposes but is also in the general interest of the radio astronomy community. 

The impact of LEO objects on low-frequency observations can also be mitigated using the SDA understanding of the non-coherent system. Based on the knowledge of FM-reflecting LEO objects predicted to be within the FOV during an observation (knowledge from maintaining a catalog of FM-reflecting LEO objects), the observation parameters could be optimised (if feasible), such as UV-weighting and correlator time-averaging, to emphasise the near-field effect and fringe-washing effect. A catalog of FM reflecting LEO objects can also be used to better understand the expected false positive rates in transient and variability studies, analogous to similar studies at optical wavelengths \citep{2021AJ....161..135A, 2021PASA...38....1T, 2020PASJ...72....3R}. 

\section{CONCLUSION}
\label{sec:conclusion}

The rapid increase in LEO objects within the last decade demands the development of orbit determination capabilities using multiple SDA sensors, to assist in avoiding cascading collision event scenarios \citep{Kessler2010KesslerPaper}. In this work, a orbit estimation capability using MWA SDA observations is demonstrated using angles-only measurements of satellite passes. 


This paper demonstrates a preliminary orbit determination capability using a wide field-of-view system, the MWA, in non-coherent passive radar mode at FM frequencies. The non-coherent passive radar techniques developed for the MWA \citep{Tingay2013OnFeasibility, zhang2018limits, prabu_dev, prabu_survey} were used in this work and the detected signals are spatially smeared over several degrees, mainly due to the motion of the satellites during the $2$\,s difference image timescales. Through Gaussian modeling of the observed satellite signals in MWA data, the work demonstrates a technique for obtaining satellite position measurements from MWA SDA observations, by understanding the different factors (such as the PSF structure and apparent satellite motion) that contribute to the smearing of the signals, and the method is tested on 32 LEO satellite detections obtained from the previous work \citep{prabu_survey}.

The angular position measurements from the method were used to perform orbit determination for the objects using a least-squares fit approach and an SGP4 propagator, and these orbital elements are reported in this paper. The orbital elements were verified against the publicly released TLE values by the Space Surveillance Network (SSN) \footnote{Obtained from \url{space-track.org}.} and were found to be in good agreement.

In the future, the angles-only position measurements from the non-coherent detection method can be coupled with measurements from other sensors to perform catalog maintenance and conjunction monitoring through data fusion. Due to fitting for the six parameters using measurements in 2D space (RA-DEC), the errors in the orbit determination process were found to be highly coupled, and range measurements (that have three independent Euclidean x,y,z components) can 
help de-couple these errors and constrain the orbital elements better. Awareness of the errors associated with the instrument is especially important when performing data fusion for joint orbital element estimation. Studies have shown data fusion to be more effective when using 3D measurements (angular position measurements with range measurements) compared to using  2D measurements (angular measurements only) \citep{bennett2015analysis}, and using range measurements
can help de-couple the uncertainties in the orbital element estimates.

Objects were successfully re-acquired in a future orbit using the determined orbital elements. The offset between the predicted pass and the observed pass was found to be reduced when using multi-pass observations to perform orbit determination, compared to a single-pass orbit determination.

The paper concludes by placing the developed MWA SDA capability in the global context of SDA sensors, by discussing the accuracy of angular position measurements, the goodness of orbit determination, and the value added to the existing global SDA network by the MWA SDA system. Based on the understanding of MWA's response towards LEO objects, methods to mitigate the impact of LEO objects on radio-astronomy observations are discussed, and how maintaining a catalog of FM reflecting LEO objects can help better understand FM band observations performed from the MRO, home to the MWA and the future low-frequency SKA.

\begin{acknowledgements}
This scientific work makes use of the Murchison Radio-astronomy Observatory, operated by CSIRO. We acknowledge the Wajarri Yamatji people as the traditional owners of the Observatory site. Support for the operation of the MWA is provided by the Australian Government (NCRIS), under a contract to Curtin University administered by Astronomy Australia Limited. We acknowledge the Pawsey Supercomputing Centre which is supported by the Western Australian and Australian Governments. Steve Prabu would like to thank Innovation Central Perth, a collaboration of Cisco, Curtin University,  Woodside and CSIRO’s Data61, for their scholarship.

\subsection*{Software}
We acknowledge the work and the support of the developers of the following Python packages:
{\tt Astropy} \citep{theastropycollaboration_astropy_2013,astropycollaboration_astropy_2018}, {\tt Numpy} \citep{vanderwalt_numpy_2011}, {\tt Scipy}  \citep{jones_scipy_2001}, {\tt matplotlib} \citep{Hunter:2007}, {\tt SkyField}\footnote{\url{https://rhodesmill.org/skyfield/}}. The work also used {\tt WSCLEAN} \citep{offringa-wsclean-2014,offringa-wsclean-2017} for making fits images, {\tt calibrate} \citep{offringa-2016} for calibrating MWA observations, and {\tt DS9}\footnote{\href{http://ds9.si.edu/site/Home.html}{ds9.si.edu/site/Home.html}} for visualization purposes. 

\end{acknowledgements}






\clearpage
\printbibliography

\end{document}